# Forensic Analysis of TomTom Navigation Application

Nhien-An Le-Khac, Mark Roeloffs and M-Tahar Kechadi

University College Dublin, Ireland

Email: {an.lekhac,mark.roeloffs,tahar.kechadi}@ucd.ie

**Abstract**

In the forensic field of digital technology, there has been a great deal of investigation into the decoding of navigation systems of the brand TomTom. As TomTom is the market leader in navigation systems, a large number of these devices are investigated. These devices can hold an abundance of significant location information. Currently, it is possible with the use of multiple methods to make physical copies of mobile devices running Android. The next great forensic problem is all the various programs that can be installed on these devices. There is now an application available from the company TomTom in the Google Play Store. This application mimics a navigation system on your Android mobile device. Indeed, the TomTom application on Android can hold a great deal of information. In this paper, we present a process of forensic acquisition and analysis of the TomTom Android application. We focus on the following questions: Is there a possibility to find previously driven routes or GPS coordinates with timestamps in the memory of the mobile device? To investigate what is stored in these files, driving tests were performed. During these driving tests a copy was made of the most important file using a self-written program. The significant files were found and the data in these files was decoded. We show and analyse our results with Samsung mobile devices. We compare also these results with forensic acquisition from TomTom GPS devices.

# 1. Introduction

In the past two decades, Global Positioning Systems (GPS) are widely used from military to commercial users around the world. Today, GPS navigation systems are becoming increasingly common. They can help motorists, pilots, sailors and the military not only to determine the device's location but also with important information such as maps, streets maps, directions, roads and their status, location of amenities (food, fuel, etc.). As a consequence, the forensic acquisition and analysis of these devices is of great interest as they have the potential of storing historic spatial data. Among GPS navigation systems, TomTom Portable Navigation Device (Tomtom PND) is one of the market leaders in Europe with a market share of almost 50% in 2008 [1]. Research on the forensic acquisition and analysis of Tomtom PND can be found in [2] and [3].

Today, smart-phones are used for all kinds of things, from calling, browsing on Internet to games. But more interesting is that users carry their mobile phones with them the whole day [4]. They browse the Internet at various locations; keep their agenda, etc. Another purpose of these smart-phones is their use as a navigation system. This is possible due to an embedded GPS-chip in most smartphones. From a forensic perspective this feature is very important. With all the data contained in a mobile device, it is possible to know its geographical location at a given time. The main challenge for mobile device forensics at present is that there is huge number of very diverse apps (applications) that are available in those devices. These applications can be installed easily at any time. Making physical copies of mobile devices, especially Android mobile devices, is no longer the problem, but the decoding of all of the data on mobile devices is the main challenge. Moreover, another crucial problem that is arising is the encryption of the complete non-volatile memory of a device. In an iPhone this is already standard, and is an option in Android mobile devices.

Currently, Android is the top smartphone platform used in the U.S [5] and one of the most important navigation system application on Android mobile device is the TomTom Android Application. It was launched in October 2012 in Google Play Store and has been downloaded thousands of times. The application is able to store multiple geographical points to which used during the navigation, and even time stamps may be available. As this application can potentially store all the important information, it can be very useful in an investigation. Besides, there is no software on the market that can decode the data from the TomTom Android Application (TomTom AA). Indeed, very few researches in literature focus on the forensic acquisition of TomTom AA.

In this paper, we present a process of forensic acquisition and analysys of TomTom AA. We show step-by-step the operation of retrieving forensic data from TomTom AA. By performing driving tests with different mobile devices the significance of the data was uncovered. The main files were found and decoded in a human-readable format, but the most important data like the GPS coordinates are stored in a special format. This format needs to be converted into human readable for analysis. We describe and analyse our results with Samsung mobile phone. We also compare the differences with TomTom PND and TomTom AA.

The rest of this paper is organised as follows: Section 2 deals with background research in the area. Section 3 presents our adopted approach for retrieving forensic data from TomTom AA. We show test results on different mobile devices and analyse and discuss the experimental results of these approaches in Section 4. Finally, we conclude in Section 5.

## 2. Related work

We firstly review some existing software used in the mobile phones forensic such as UFED, XRY and TomTology. We then discuss related research of analysing TomTom PNDs and TomTom AA.

UFED Physical Analyser (PA) is software/hardware made by Cellebrite Ltd. The hardware is the UFED Touch [6], a standalone computer with a touch screen running Windows. This computer is able to read a great number of mobile phones, both logically and physically. The extracted data can be stored on a computer, a USB stick/hard drive or SD card. The software on the UFED Touch cannot decode a physical image. UFED Physical Analyser is used for this purpose. For all decoded data with an image of a mobile device is fairly easy. Moreover, UFED Touch is equipped to perform advanced search. There are ways of searching, for example, for 7-bit encoded text from an SMS message. All the decoded data can be found in the memory image; UFED PA will show the data in a colour in the image. There is also Python scripting language support within UFED PA, which will enable a user to decode data, which UFED PA cannot find. XRY Physical is software/hardware made by Microsystemation A.B. (MSAB) [7]. The hardware consists of a USB Hub with a Bluetooth USB and USB Infrared device. There is a field version with a ruggedized device, but XRY does not have a dedicated device like the UFED. The software is able to make logical and physical images for a number of models of mobile phones. The decoding of dumps is comparable to UFED PA, but finding the data back in a memory copy is not as extensive as in UFED PA. XRY also has a Python interface. TomTology is software developed by Forensic Navigation Ltd [8]. TomTology is a software tool to decode memory images made of TomTom devices. TomTology is able to decode TomTom images from TomTom PNDs from the first version. Because there is no way to make a memory dump of a TomTom PND of the second series via the cable, there is no method to decode the TomTom PNDs of the second series. The forensic software on the market like UFED PA and XRY physical are able to make images of Android mobile phones. The file system of these memory images can be decoded for most Android mobile devices. The next problem is the decoding of applications. At present there is no way to decode the TomTom AA with UFED PA or XRY Physical.

Also TomTology is not able to decode images from Android devices; it will only decode TomTom images made from TomTom PNDs of the first series. To the best of our knowledge, there is no software available to decode information of the TomTom AA.

There is very few research on analysing TomTom devices in the literature and most of the research focus on the TomTom PNDs not TomTom AAs. For a mobile device as used in [9] it is slightly harder. First the mobile device has to be rooted to be able to make a copy. The program *dd* is used to make this copy. Next, there is a copy of the whole non-volatile memory in a mobile device. In Nutter [10], it first extracts a physical image to retrieve the most important files. The files have then to be decoded. The file mapsettings.cfg in the mobile devices has not been found, but there is a similar file decoded [10]. Also there are multiple files found in the unallocated clusters. In some of the mobile devices that were analysed in [10] there is also data in these unallocated clusters, but for one device there is also no deleted data. Other related work on the mobile phone forensics can be found in [15][16], they do not however focus on TomTom AAs. To the best of our knowledge, there is no full research on the acquisition and analyse TomTom AAs.

## 3. Acquisition Techniques

The proposed acquisition technique consists of three stages: physical image extraction, locating important files, and finally decoding/analysing these important files. In the following we describe each of these stages.

### *3.1. Physical Image Extraction*

To make a physical image of the non-volatile memory of an Android mobile device, one has to have root access to the device. *Bootloader* methods for making memory copies of multiple mobile devices are usually complex to use. Moreover, there are multiple ways of getting root access to all sorts of Android mobile devices. The first step is to find out how a given mobile device can be rooted, as different mobile devices have different versions of Android, so the methods for gaining root access can be different. These methods are well documented on the web.

### *3.2. Important files searching*

The files found in a mobile device are not all forensically useful, so once root access has been granted the crucial task is to identify the interesting files. We can always start by getting favourite data, address or location, but usually, the main task is to make physical copy of the whole device. By comparing the two memory copies (before and after entering testing data), the interesting files can be identified and extracted. Because one can experience a lot of changes between the memory copies, it is easier to perform the comparison between the file systems of the mobile phones. The process of searching for useful files follows these steps: (i) Load the memory copy in software that will decode the file system like UFED Physical Analyzer; (ii) Search for the TomTom folder by searching for 'tomtom'; (iii) Look at the files in the TomTom folder to see if the files are there.

### *3.3. Decoding the Files*

The identified files were decoded and then investigated to determine, for instance, how the data is stored within those files. Because the information is stored in xml format, there is also meta-data available [11]. We wrote scripts to decode those files and extracted very interesting and relevant information. The outputs of these scripts are presented as tables in order to easily interpret them.

### 4. Description of results and discussion

### *4.1. Samsung mobile device acquisition*

One of the most popular Samsung Android smart-phones in the Netherlands, at the start of this research, is the Samsung Galaxy S3 (GT-i9300). Therefore this is the main mobile device used for this research and the mobile device was updated to the most recent version.

As previously stated, to conduct the research it was necessary to make multiple physical images after each round of testing. It is possible to make a physical image using the Cellebrite UFED Touch, but there was no UFED Touch available to the researcher on a daily basis. Therefore another method was used to make physical images of the mobile device. The memory of a Samsung Galaxy S3 is a 16GB eMMC chip, the Samsung KMVTU000LM [12].

## *4.2. Decoding the important files*

### 4.2.1. Voices

The TomTom Android program has a *VoiceProvidersDatabase* database that contains the locations for the different voices. For a number of languages, the TomTom Android Application has built-in Voices. The audio files containing the voices, which tell you for example to turn right or turn left, are stored in the external microSD card (if it is available) or in the internal mass storage device. If there is no large user area, the application will not work. The first time the program starts; the voices and chart material are downloaded and stored. The voice information has not been further investigated, as this does not change between routes.

### 4.2.2 Favourites

The favourites are stored in the file called *Favorites.ov2*. The *ov2* file is also available on the TomTom Portable Navigation Devices. The specification of the *ov2* format has been published in [13]. The header of an *ov2* file is shown in Figure 1.

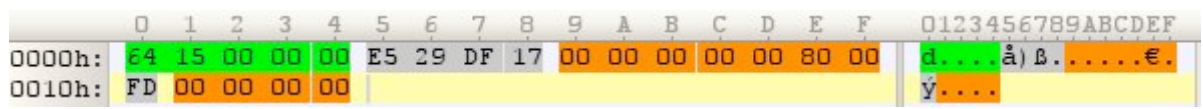

Figure 1: ov2 file header

In this Figure, the first five bytes shown in green is the *ov2* file header. The bytes in grey are magic numbers specific to *ov2* files; these are different according to the various models of Android devices. The bytes in orange colour are the same for all *ov2* files.

```
        0  1  2  3  4  5  6  7  8  9  A  B  C  D  E  F   0123456789ABCDEF
0000h:  02 35 00 00 00 C2 33 07 00 4F 60 4F 00 52 69 64   .5...Â3..O`O.Rid
0010h:  64 65 72 20 44 69 72 6B 73 74 72 61 61 74 20 2D   der Dirkstraat -
0020h:  20 53 6F 70 68 69 61 73 74 72 61 61 74 2C 20 47    Sophiastraat, G
0030h:  6F 75 64 61 00                                    ouda.
```

Figure 2: *ov2* favourite entry

Moreover, an entry of a favourite is shown in Figure 2. The byte in yellow indicates that it is a current favourite. After the first byte, the bytes in green denote the length (in this case 0x35) of this entry. The next 4 bytes in red denote the longitude and then the 4-byte latitude in orange. The user string is the last part on the entry, which is a null-terminated ASCII string, coloured in blue. Note that the home location is not stored as a Favourite. So, the regular expression for seeking a *Favourite* file would be:

**\x64\x15\x00{3}[\x01-\xFE]{4}\x00{5}\x80\x00[\x01-\xFE]{1}\x00{4}**

Besides, we can have multiple programs that can interpret the *ov2* file. One of these programs is Poiedit (www.poiedit.com).

### 4.2.3 Unique ID

Another important file is Benelux file (Benelux_XXXXXXXX.xml). The XXXXXXXX represents a number that is specific for a particular mobile device. Each model gets such a (non-unique) number. For example, the Samsung Galaxy S3 is Benelux_AF7DE92B.xml. This is an xml file as shown in the following Figure 3.

```
<string name="MapSettings*00000*/AddressRecents*00000*/AddressRecents_Address*00023*/Location_Line*00023*/LineRec_MaxSpeed*00023*">MA==</string>
<string name="MapSettings*00000*/AddressRecents*00000*/AddressRecents_Address*00021*/Location_NodeFrom*00021*/NodeRec_Delta*00021*">NTEy</string>
<string name="MapSettings*00000*/EngineRecents*00000*/EngineRecents_Recent*00003*/Location_NodeUpto*00003*/NodeRec_Pos*00003*">KDQ3MTMwODsgNTIwMTgxNCk=</string>
<string name="MapSettings*00000*/AddressRecents*00000*/AddressRecents_Address*00025*/Location_Line*00025*/LineRec_Type*00025*">Mzk=</string>
<string name="MapSettings*00000*/EngineRecents*00000*/EngineRecents_Recent*00000*/Location_Line*00000*/LineRec_NamePos*00000*">MjYxMTU3</string>
<string name="MapSettings*00000*/NeverAskedDefaultCountry*00000*">ZmFsc2U=</string>
<string name="MapSettings*00000*/AddressRecents*00000*/AddressRecents_Address*00002*/Location_Line*00002*/LineRec_MaxSpeed*00002*">MzA=</string>
<string name="MapSettings*00000*/SafetyCameraWarnings*00000*/SafetyCameraWarnings_Warning*00007*/SafetyCameraWarnings_Warning_WarningDistance*00007*">NTAwMA==</string>
<string name="MapSettings*00000*/AddressRecents*00000*/AddressRecents_Address*00013*/Location_PoiType*00013*">LTE=</string>
```

Figure. 3: Part of a Benelux_XXXXXXXX file

As shown in Figure 3, the most promising part of this xml file is the data immediately preceding the final </string> tag. This is the data, which is formatted in such a way that it is not human readable. Generally strings are stored in ASCII format as seen in the rest of the xml file. As the string ends multiple times in ==, it was concluded that the strings are of base 64 encoded [14]. When the data of base 64 is decoded, the navigated street names become visible. Base 64 is used in this application so as to support foreign characters. Indeed, the data in this xml is not sorted i.e. data from an entered address is stored in multiple lines not necessarily consecutive, but appear throughout the entire file. We also wrote scripts to sort out the data and decode the base 64 data. In the file generated from these scripts, the following main xml items were recovered: *EngineRecents; AddressRecents; NeverAskedDefaultCountry; SafetyCameraWarnings; LastSelectedPoi; PoiSet; RouteStream; LastSelectedSearchItem; LastKnownTrueGpsPosX; LastKnownTrueGpsPosY; PDKAutoShutdown; PDKDisableiPodMenuIcon; RegularRouteLocHome; RegularRouteLocWork; LastSelectedPoiData; PDKUseDefaultSettingsFromUserFile; TrafficOnMap; SoundVolumeHandsFree; TrafficAutoUpdate; TrafficAutoReplan; TaiwanCenterAvailable; CurrentTrafficRouteType; PoiWarnings; GeoFormat; CurrentSelectedCountryIndex; TrafficUpdateFrequency; UserEnabledTraffic; PoicatHotlistCat; SupportASN; ValidPassword; PoicatHotListValidatedOnce; PoicatHotlistHit; UserMarkerAvailable; TrafficWarnings; EnableBT*.

Among these items above, the most interesting and relevant data for forensic purposes can be listed as follows:

• **EngineRecents**: contains the locations, which are recently visited. These locations are not only addresses, but may also be, for example, Points Of Interest (POI) or a particular point on the map. These are the recent destinations that TomTology decodes in the TomTom PNDs. Datatype EngineRecents contains a number of subtypes, most of which are utilised for the calculation of routes. The most relevant subtypes are:

    (i) *Location_UserName*: The name the user gives to a location. This is not the case of an address, but in the case of a Favourite location to which the user may give a name.

(ii) *Location_UserPos*: The position of the location in a specified coordinate system, usually specified in decimal degrees. This is the point the user enters into the device.

(iii) Location_LocName: In the case of a street address, the street is saved in this data type. If the location is a city, the city name is stored here.

(iv) Location_LocType: The types of location, the following types have been identified:

| | |
|---|---|
| LOCTYP_MAPTICK | Navigated to a "Point on map" or to a "Latitude Longitude" |
| LOCTYP_ADDRESS | Navigated to an "Address" or a Contact |
| LOCTYP_HOME | Navigated to "Home" |
| LOCTYP_POI | Navigated to a "Point of Interest" |
| LOCTYP_undefined | No defined location asyet |
| LOCTYP_FAVOURITE | Navigated to a Favourite |
| LOCTYP_GPS | The current GPS location. If this is in a RouteStream it can also be the last known position. This current GPS position may be incorrect if there was no GPS coverage, and so the last known GPS location is stored. |

(v) Location_CityName: The name of the city where the location is situated.

(vi) HouseNumber_Number: The house number of the address.

There is no option to see if a location was visited, except if the Location_LocType is LOCTYP_GPS, then the location was visited at some point in time.

- **AddressRecents**: These are the recent entered addresses. In the EngineRecents are the locations navigated to, which are not addresses, are also stored. In the AddressRecents data type only addresses are stored. The sub-data types are the same as in the EngineRecents data type.

• ***RouteStream*:** The RouteStream holds the departure location, the destination location and the departure time. If the device has a GPS lock, the departure location is the current GPS location. If not, the last known location is stored here. The departure time is dependent on the clock of the device, not of GPS-time. So if the time of the device is set incorrectly, the departure time is also wrong. If the time is set correctly and the device has a GPS lock, the departure location was visited at the departure time.

• ***LastSelectedPoi***: The LastSelectedPoi holds the last selected point of interest (POI). The sub-data types are the same as in the EngineRecents data type.

• ***LastSelectedSearchItem*:** In the TomTom application there is an option to do a Local Search. With this local search TomTom searches for a location near you. For example, if you want to go to a clothing shop, you can enter the name of the shop and the TomTom will try to find it using an Internet connection. In the local search screen, the last selected search item can be seen. This item is stored with the type LastSelectedSearchItem. The sub-data types are the same as the ones in the EngineRecents data type.

• ***RegularRouteLocHome/Work***: During this research, there was no option to fill this data type. The name suggests this is an option to store regular route, so that it is easy to plan. It could be used to see the amount of traffic on the home/work route. The sub-data types are the same as the ones in the EngineRecents data type.

• ***LastKnownTrueGpsPosX/PosY*** store the last known GPS position. There is no GPS time stored.

• ***LastSelectedPoiData***: The data of the last selected POI. An interesting point is that the GPS location of this data type first needed to be halved in order to get the same location format as the rest of the locations.

### 4.2.4 NavkitSettings.xml

As a next step, we analyse a significant file: NavkitSettings.xml. The home locations are not stored in the Benelux_XXXXXXXX, but in the NavkitSettings.xml, and this file holds a range of settings too. It is also an xml-formatted file with base64-coded data. This file contains a great deal of data types, for a range of settings. Not all of these data types are interesting in this research:

• *UP_HomeLocations*: This data type stores the home locations. There can be multiple home locations stored in the TomTom application. There are numbers associated to the locations. The location with the highest number is the current home location. The sub-data types are the same as the ones in the EngineRecents data type.

• *TTPlusManager* stores all the paid subscriptions that are activated in the TomTom application. There can be subscriptions for Mobile HD traffic, TomTom Places, Free POIs, Free Maps and Free Voices. With these subscriptions, a start and end time is stored. In the TTPlusManager data type there is also an entry for the username, password, ConnectionData_LastValidTime, ConnectionData_LastConnectionTime and AccountInfo_DatelastUpdate. The times for ConnectionData_LastValidTime, ConnectionData_LastConnectionTime and AccountInfo_DatelastUpdate are off by one month and one day. This has been observed in all three mobile devices; it appears that the time on the TomTom servers is off by one month. The time is correct; it is the time in GMT (UTC).

• *LastDockedPositionY, LastDockedPositionX, LastDockedTime*: These three data types are combined, as they all store something about the last docked location. The LastDockedPositionX and LastDockedPositionY are the longitude and latitude respectively of the last docked position. The coordinates are stored in the same manner as in the EngineRecents data type. The time, however, is stored somewhat differently from the other times, which were already found. It is stored in minutes instead of seconds. So to decode the time it has to be multiplied by 60 to get to the time format used in the rest of the application.

• *MapUpdateLastReminderDate, LastMapShareConnectionReminder, LMGDisplayDate, LastMapShareSubscriptionReminder, LastTimeTempBTEnabled* are dates and times. During our research, we were not able to change the values of these data types. They cannot be used in this version of the application.

• *UserTimeOffset*: This is the offset of the clock in the mobile device in seconds. For example, when a time offset of GMT +2 (summer time in the Netherlands) is on, the value of this variable is 7259. If this value is divided by 3600 it is about 2 hours.

• *ArrivalTime*: for all the three tested mobile devices, its value is 86401. During testing it was not possible to change this value. This leads to the conclusion that this value cannot be changed by the application.

• *LocalSearchService*: In this data type, the search history is stored. This search history is the history of the local search option. If the same term is searched for two or more times, it is stored only once.

### 4.2.5 Times and Dates in the TomTom Android Application

In the TomTom PNDs there is not that much time/date information connected to a GPS coordinate. The most interesting in a TomTom PND is the *triplog* data, which contains the driven routes with date and time. It is a breadcrumb trail the TomTom PND has driven. The TomTom Android Application does not have this breadcrumb file, so a big part is not available.

There is, however, something else. There is the last docked location with the time. This is one of the most interesting points stored in the TomTom Android Application. Together with the departure time, which can be coupled to a GPS coordinate.

Unfortunately these time/date stamps are the time/date that is set in the mobile devices, so it could be wrong. Also in the case of the departure time, if the data type is LOCTYP_GPS, it could be wrong if there was no GPS lock on that place. In short, there is not that much time/date related data in the TomTom Android Application. It is very hard to identify where the mobile device was at a certain point of time with the data found in the TomTom Android Application.

We have described and analysed main files used by the TomTom Android Application. Further analysis could be carried out upon the log files outside of the TomTom application folder. This research did not address those files. Not all of the data types in certain files are decoded, as there is an excessive amount of overhead information. These data types could be analysed to determine if they also hold interesting information. More research can be put into those files. As stated earlier, there are multiple TomTom applications in the Google Play Store. The application chosen for this research was the TomTom Benelux application. It is assumed that the other TomTom applications would work in a similar manner.

## *4.3. Comparison Between TomTom PNDs & TomTom Android Application*

The devices chosen for this comparison are listed in Table 1. The TomTom Go 720 is a model of the first generation of TomTom PND systems. The main data is stored in the file mapsettings.cfg, which can be decoded with the program TomTology.

The Via 825 Live is a TomTom PND of the second generation. It is no longer possible to get a physical image of this device by USB cable. It is possible to perform a chip-off, and there is also a non-destructive method available for these sort of devices, which cannot be disclosed. The main data is stored in the mapsettings.tlv file, but there are more files, which also store data related to the data found in the first series of TomTom PNDs.

| Model | Version | |
|---|---|---|
| Go 720 | | 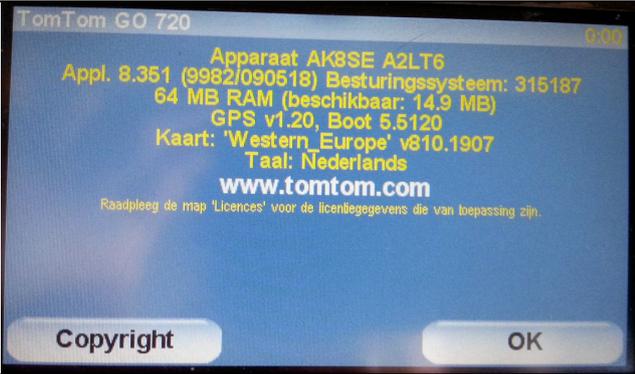 |
| Via 825 Live | | 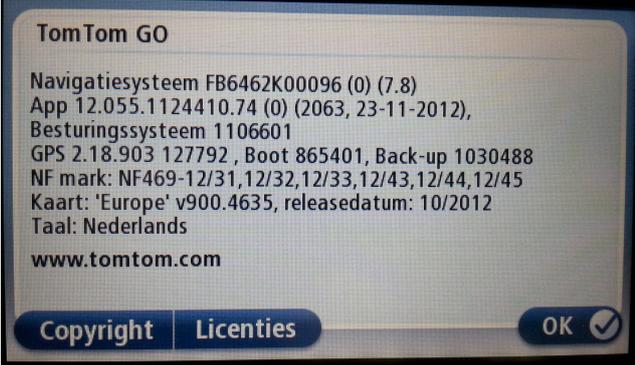 |

Table 1: TomTom PND used in experiments

Table 2 shows the various data, which can be found in different places in the PNDs of the first and second generations and the TomTom AA.

| | **TomTom PND first series** | **TomTom PND second series** | **TomTom Android Application** |
|---|---|---|---|
| *Triplogs* | Statdata folder | Statdata folder | Not Found |
| *Home Location* | mapsettings.cfg | userpatch.dat | NavkitSettings.xml |
| *Favourites* | mapsettings.cfg | Favorites.ov2 | Favorites.ov2 |
| *Recent Destinations* | mapsettings.cfg | mapsettings.tlv | Benelux_XXXXXXXX.xml |
| *Entered locations* | mapsettings.cfg | mapsettings.tlv | Benelux_XXXXXXXX.xml |
| *Journeys* | mapsettings.cfg | mapsettings.tlv | Benelux_XXXXXXXX.xml with departure time |
| *Last docked* | mapsettings.cfg | Userpatch.dat | NavkitSettings.xml, with a time stamp |
| *Bluetooth coupled devices* | mapsettings.cfg | Settings.tlv | Handled by the Android OS, not by the TomTom Android Application |
| *Simcard data* | Not applicable for first generation PND | mobility.sim | Handled by the Android OS, not by the TomTom Android Application |

Table 2: Comparison between TomTom PND and TomTom AA

## 5. Conclusion and future work

This study focuses on the investigation of mobile devices running Android operating system and its versions. The devices have 92% of market share as of May 2013. In this paper, we presented a process of forensic acquisition and analysis of the TomTom Android application on Samsung mobile devices. With UFED PA the file system is decoded to a tree views, it is possible to read all the files in the mobile devices and see what is in there. To discover the meaning of the different files in a directory some small tests were performed and the most interesting files were found. To find the meaning of the data in these files, driving tests were performed. After each driving test a copy was made of the most interesting file. We use a self-written Android Application to make copies of the device content. By analysing the results of these three tests, we notice that the way of storing data in the TomTom Android Application is different than the way the data is stored in the TomTom PNDs. In the PNDs the data is stored in a binary format, while the xml format together with base64-coded data are used in the TomTom Android Application. Both types of TomTom devices use the same sort of data, for example, one point is stored with multiple other points in the vicinity. The idea is that these points are used for the routing algorithm. However, not much time data together with GPS coordinates are found. There can be more data in the Android OS itself. In earlier versions the GPS coordinates were also stored in the Android OS itself. More investigations have to be performed on GPS applications running on the Android OS.

Also more investigation can be performed in the different payable options in the TomTom Android Application. There are the options for TomTom HD traffic and for Speed Cameras and Danger Zones. Maybe these options also provide more data to be stored by the TomTom Android Application. We are doing the acquisition and analysis of TomTom AAs installed on other mobile smartphones such as Sony Xperia. Besides, we also investigate the deleted files of TomTom AAs.